\documentstyle[twoside,psfig,epsf,amsfonts]{amsart}

\begin{document}

\title{KRIGING SCENARIO FOR CAPITAL MARKETS}

\author{T. SUS{\L}O}

\begin{abstract}
An introduction to numerical statistics.
\end{abstract}

\maketitle
 
\thispagestyle{empty}

\setcounter{footnote}{0}
\renewcommand{\thefootnote}{\alph{footnote}}

\vspace*{4pt}
\normalsize\baselineskip=13pt  
\section{Introduction}
\noindent
\label{sec:1}
\noindent Let us consider an unknown constant mean $m$
and a variance $\sigma^2$ of the random variable $V$
and the linear unbiased estimation statistics 
(weighted variable with sum of its weights equal to one) at $j \ge n+1$
$$
\hat{V}_j=\sum_{i=1}^n \omega^i_j V_i = \omega^i_j V_i
$$ 
with minimized (the best statistics)    
\begin{equation}
E\{[(V_j-m)-(\omega^i_j V_i-m)]^2\}
\label{1lhside}
\end{equation}
and minimized   
\begin{equation}
E\{[\omega_j^i V_i-m]^2\} 
\label{2lhside}
\end{equation}
our aim is to satisfy for finite $n$ 
the asymptotic property  of~(\ref{1lhside})
$$
\lim_{n \rightarrow \infty}
E\{[(V_j-m)-(\omega^i_j V_i-m)]^2\}
= 
\sigma^2  
$$
and the asymptotic property of~(\ref{2lhside})
$$
\lim_{n \rightarrow \infty}
E\{[\omega_j^i V_i-m]^2\} 
=0
$$
equivalent to
$$
\lim_{n \rightarrow \infty}
\omega^i_j V_i
=
\lim_{n \rightarrow \infty}
\omega^i_j v_i
=
m
$$
to replace (conditionally) weighted variable $\omega^i_j V_i$ by 
weighted average $\omega^i_j v_i$. 

\section{The numerical statistics}

\noindent The minimized variance of the
estimation statistics~(\ref{2lhside})  
\begin{equation}
E\{[\omega^i_j V_i-m]^2\}
=
\mp\sigma^2
\left(\sum_{i=1}^n \omega^i_j \rho_{ij} - \mu_j \right) 
=
\mp\sigma^2
\left(\omega^i_j \rho_{ij} - \mu_j \right) 
\label{vofes}
\end{equation}
and~(\ref{1lhside}) 
\begin{equation}
E\{[(V_j-m)-(\omega^i_j V_i-m)]^2\}
=
\sigma^2
\left( 1 \pm \left(\omega^i_j \rho_{ij} + \mu_j\right)\right) 
\label{voffue}
\end{equation}
in terms of correlation function $\rho_{ij}$
is given by the kriging system of equations
\begin{equation}
\begin{array}{cccccl}
{\underbrace{
\left[
\begin{array}{cccc}
\rho_{11} & \ldots & \rho_{1n} &  1 \\
\vdots  & \ddots & \vdots & \vdots \\
\rho_{n1}&  \ldots & \rho_{nn} &  1 \\
1  & \ldots & 1 & 0 \\
\end{array}
\right]}_{(n+1)\times(n+1)}}
& 
\cdot 
&
\underbrace{
\left[
\begin{array}{c}
\omega_j^1 \\
\vdots  \\
\omega_j^n \\
\mu_j \\
\end{array}
\right]
}_{(n+1)\times 1}
&
=
&
\underbrace{
\left[
\begin{array}{c}
\rho_{1j} \\
\vdots  \\
\rho_{nj} \\
1 \\
\end{array}
\right] 
}_{(n+1) \times 1} 
\end{array} \ . 
\label{keq}
\end{equation}
To replace (conditionally) weighted variable by 
weighted average in~(\ref{vofes}) 
\begin{equation}
E\{[\omega^i_j V_i-m]^2\}
=
\min[\omega_j^i v_i -m]^2 
=
\mp\sigma^2
\left(\omega^i_j \rho_{ij}-\mu_j \right)   
\label{mse}
\end{equation}
we have to find (on computer) a numerical disjunction of 
the best linear estimation statistics
which satisfies the asymptotic property of~(\ref{voffue})  
constrained by numerical approximation 
to root of the equation
\begin{equation}
\omega^i_j \rho_{ij} + \mu_j = 0 
\label{lsc}
\end{equation}  
and fulfills the asymptotic property of~(\ref{vofes})
for $j \rightarrow \infty$.

\vspace*{12pt}
\noindent
\noindent {\it Proof.} Since for $j \rightarrow \infty$  
the correlation has to vanish   
$$
\underbrace{
\left[
\begin{array}{c}
\rho_{1j} \\
\vdots  \\
\rho_{nj} \\
\end{array}
\right] 
}_{n \times 1}
=
\underbrace{
\left[
\begin{array}{c}
\rho(|1-j|) \\
\vdots  \\
\rho(|n-j|) \\
\end{array}
\right] 
}_{n \times 1}
=
\xi
\underbrace{
\left[
\begin{array}{c}
1 \\
\vdots  \\
1 \\
\end{array}
\right] 
}_{n \times 1}
\qquad \xi \rightarrow 0^- ~(\mbox{or} ~\xi \rightarrow 0^+ )
$$ 
the constraint~(\ref{lsc}) can be easy fulfilled 
$$
\omega^i_j \rho_{ij} + \mu_j 
= 
\xi \underbrace{(\omega^1_j+\ldots+\omega^n_j)}_{1} + \mu_j 
= 
\xi + \mu_j 
= 
0 
$$ 
the kriging system that does not depend on $j$ now
\begin{equation}
\begin{array}{cccccl}
{\underbrace{
\left[
\begin{array}{cccc}
\rho_{11} & \ldots & \rho_{1n} & 1 \\
\vdots  & \ddots & \vdots & \vdots \\
\rho_{n1}&  \ldots & \rho_{nn} & 1 \\
1  & \ldots & 1 & 0 \\
\end{array}
\right]}_{(n+1)\times(n+1)}}
& 
\cdot 
&
\underbrace{
\left[
\begin{array}{c}
\omega^1 \\
\vdots  \\
\omega^n \\
- \xi \\
\end{array}
\right]
}_{(n+1)\times 1}
&
=
&
\underbrace{
\left[
\begin{array}{c}
\xi \\
\vdots  \\
\xi \\
1 \\
\end{array}
\right] 
}_{(n+1) \times 1}
&
\end{array}
\label{skeq}
\end{equation}
or (if rewritten) 
$$
\Lambda\omega-\xi F=\xi F 
$$
and 
$$
F'\omega=1 \ ,
$$
where
$$
\begin{array}{ccccccccccc}
\omega
&
=
&
\underbrace{
\left[
\begin{array}{c}
\omega^1 \\
\vdots  \\
\omega^n \\
\end{array}
\right]
}_{n\times 1} \ ,
&
F
&
=
&
\underbrace{
\left[
\begin{array}{c}
1 \\
\vdots  \\
1 \\
\end{array}
\right] 
}_{n \times 1} \ ,
&
\Lambda
&
=
&
\underbrace{
\left[
\begin{array}{ccc}
\rho_{11} & \ldots & \rho_{1n}  \\
\vdots  & \ddots & \vdots  \\
\rho_{n1}&  \ldots & \rho_{nn}  \\
\end{array}
\right]}_{n \times n}    
&
=
&
\Lambda' \ ,
\end{array}
$$
gives the classic least-squares weights for 
an estimate $\omega'{\bf v}$  
$$
\omega=2\xi\Lambda^{-1}F
=
\left(F'\Lambda^{-1}F\right)^{-1} \Lambda^{-1}F
=
\frac{\Lambda^{-1}F}{F'\Lambda^{-1}F} \ ,  
$$
where from~(\ref{mse}) constrained by~(\ref{lsc}) holds
$$
\min[\omega'{\bf v} - m]^2  
=
\pm\sigma^2 2 \mu_j  
=
\mp \sigma^2 2 \xi 
\qquad \xi \rightarrow 0^- ~(\mbox{or} ~\xi \rightarrow 0^+ )
$$
based on
$$
{\bf v}
=
\underbrace{
\left[
\begin{array}{c}
v_1 \\
\vdots  \\
v_n \\
\end{array}
\right] 
}_{n \times 1} \ .
$$

\vspace*{12pt}
\noindent
{\bf Example.} The numerical least-squares estimator constrained 
by~(\ref{lsc}) which also fulfills the asymptotic property of~(\ref{vofes}) 
is derived by frozen model (see~Tab.~\ref{Tab1}).  
\begin{table}[h!]
\centerline{
\begin{tabular}{|c|c|c|c|c|c|c|}
\hline
Fig. & Index  & n & j & $\omega^i_j v_i$ & 
$\min[\omega_j^i v_i -m]^2$ & $\omega^i_j\rho_{ij} + \mu_j$   \\ 
\hline
1 & FTSE 100  & 132 & 426 & 8463 & 0.003150 & -0.000058   \\
\hline
2 & DJI & 113 & 356 & 13604 & 0.002569 & 0.000007  \\
\hline
3 & S\&P 500 & 102 & 270 & 1789 & 0.033930 & 0.000252 \\   
\hline            
\end{tabular}}
\caption{\label{Tab1}\small 
The numerical least-squares estimator $\omega^i_j v_i$ of 
final mean of frozen model tries at some $j$ for input sample of 
size $n$ to fulfill the asymptotic property of the best linear unbiased 
statistics.}
\end{table}
\begin{figure}[b!] 
\vspace*{13pt}
\centerline{
\psfig{file=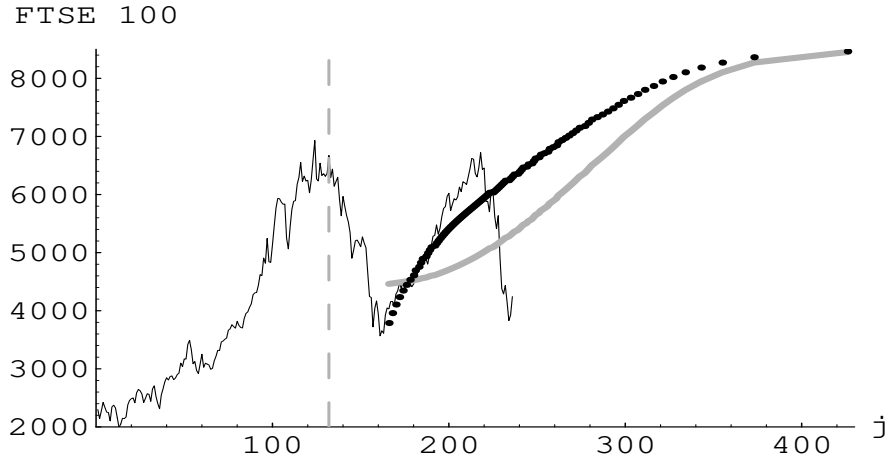,width=12cm,height=6cm}
} 
\vspace*{13pt}
\caption{\label{Fig1}\small 
FTSE 100 from 1 September 1989 up to 1 May 2009 (237 monthly close quotes). 
The classic (grey line) least-squares estimator of mean compared for frozen 
model with the numerical (black dots) least-squares estimator of mean at its 
co-ordinate. 
The assumption $j \rightarrow \infty$ is fulfilled 
at $j=426$. Dashed vertical line denotes $n=132$.
} 
\end{figure}
\begin{figure}[t] 
\vspace*{13pt}
\centerline{
\psfig{file=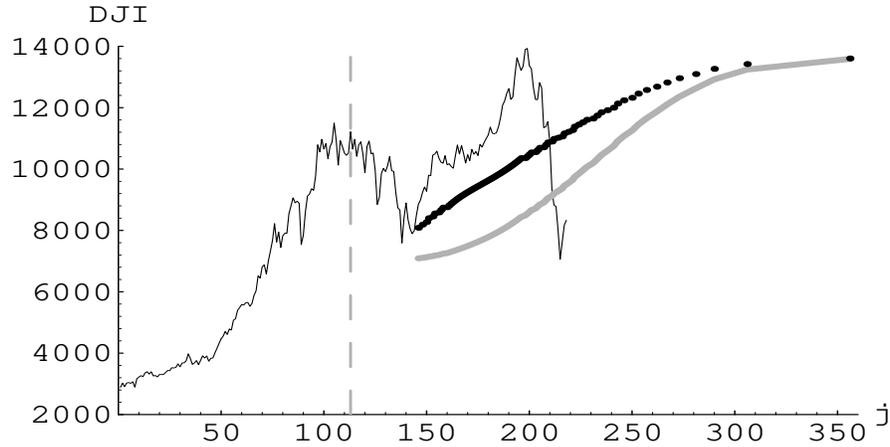,width=12cm,height=6cm}
} 
\vspace*{13pt}
\caption{\label{Fig2}\small 
DJI from 1 April 1991 up to 1 May 2009 (218 monthly close quotes). 
The classic (grey line) and numerical (black dots) least-squares estimator 
of mean for frozen model. The assumption $j \rightarrow \infty$ is fulfilled 
at $j=356$. Dashed vertical line denotes $n=113$.
} 
\end{figure}
\begin{figure}[b] 
\vspace*{13pt}
\centerline{
\psfig{file=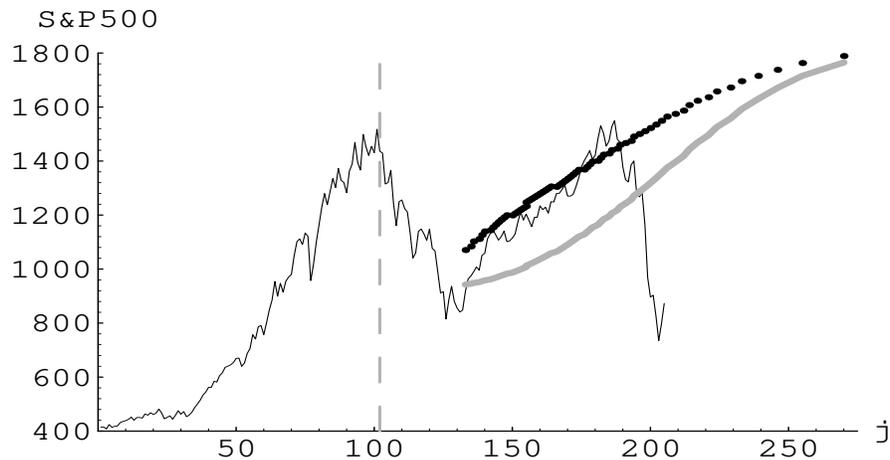,width=12cm,height=6cm}
} 
\vspace*{13pt}
\caption{\label{Fig3}\small 
S\&P 500 from 3 April 1992 up to 1 May 2009 (206 monthly close quotes). 
The classic (grey line) and numerical (black dots) least-squares estimator 
of mean for frozen model. The assumption $j \rightarrow \infty$ is fulfilled 
at $j=270$. Dashed vertical line denotes $n=102$.
} 
\end{figure}

\vspace*{12pt}
\noindent
{\bf Remark.}
Crisis emerges always when someone's wish is to
hide the cash in pocket instead of spending it. This is also the 
nature of present crisis. The root of capitalist production 
consists in industrial production of high quality components and 
semi-manufactured goods for the needs of world economy. Each 
dollar spent in any place of the world will sooner or later 
reinforce the roots of capitalism. In situation of the lack of 
such sound root, the effects of positive balance of payments can 
be deposited only in foreign exchange reserves, to avoid immediate 
consumption. The money that cannot be spent can be called dead 
money. Global financial crisis has its cause in the outflow of 
money from the international financial network, towards the 
monetary reserves of the systems which, having no sound capitalist 
roots, while enjoying high share in global trade, froze the assets 
in reserves. 
Under the impact of the crisis, the foreign exchange reserves 
held by above mentioned systems have started to shrink. Sooner or 
later, they will feed the roots of capitalism. The revitalized and 
released dead money will result in growth of inflation 
in industrial countries. This development will herald the end of 
financial crisis and the beginning of a new economic boom, since 
investors will start to buy up the shares in order to gain 
protection against inflation.

\end{document}